\documentclass{article}

\title{The One Page Model Checker}
\author{Jason E. Holt $\langle isrl@lunkwill.org \rangle$}

\begin{document}
\maketitle

\begin{abstract}
We show how standard IPC mechanisms can be used with the fork() system
call to perform explicit state model checking on all interleavings of
a multithreaded application.  We specifically show how to check for
deadlock and race conditions in programs with two threads.  Our
techniques are easy to apply to other languages, and require only the
most rudimentary parsing of the target language.  Our fundamental
system fits in one page of C code.
\end{abstract}

\section{Introduction and Related Work}

Debugging multithreaded applications is hard.  Race conditions mean
that failures may be nondeterministic.  Deadlock can be hard to trace,
because it involves behaviors from multiple concurrent threads.  Tools
to prove that a piece of code has no such behaviors can help find such
errors, and instill confidence that the programs will work correctly
when deployed.

Here we describe a method for measuring the behavior of multithreaded
programs through all possible execution interleavings.  Our work is
straightforward and applicable to many different programming
languages, although it also has some significant, fundamental
limitations.

Several other techniques have been proposed which relate to model
checking multithreaded applications.  Visser et al\cite{visser} built
an optimized system which implements the Java VM and can prove
properties about multithreaded applications.  Mercer and
Jones\cite{estes} built an explicit state model checker designed
around specific CPUs which verifies properties of compiled code.

In 1997, Savage et al \cite{eraser} introduced a tool which enforces a
locking discipline on resources to prevent nondeterministic behavior
caused by OS scheduling.  While not a model checker per se, this tool
also aims to help programmers reduce the uncertainty associated with
multithreaded applications.

\section{System Overview}

The principle behind our system is easy to understand.  Given a
program with two threads, we wish to search for particular conditions
like deadlock among all possible thread interleavings.  For example,
if the first thread executes $print(ab)$ and the second executes
$print(12)$, then at least the following behaviors are possible:
the first thread could run to completion, followed by the second,
producing the output string ``ab12'', or the second could run to
completion first, producing ``12ab''.  If the operating system chooses
to switch between the threads while they're running, the strings
``a1b2'', ``a12b'', or ``1a2b'' might also occur.  Of course, if
$print()$ isn't thread safe, the program might also exhibit other
behaviors or crash entirely.

To test all possible interleavings, the model checker must have a way
of trying different execution paths and must keep track of which paths
have been explored.  Our system's simplicity and compactness comes
from using the Unix-standard $fork()$ system call, which forks a
process at the point of the function call into two independent,
identical processes.  $fork()$ can be called inside branching, looping
or subroutine constructs, and both the original and newly created
child processes will return to the following statement and continue as
if nothing happened.  This is different from most thread
implementations, which must be called with a subroutine to execute in
the new thread, after which the thread terminates.

$fork()$ allows our model checker to be implemented much like a normal
recursive depth-first search.  In such a search, the program's stack
is used to implicitly keep track of the current progress through the
state space being searched.  In place of nested function calls, our
implementation creates a child process at each branching point to
explore the next level of the search.  This can happen surprisingly
quickly, due to efficient OS techniques like copy-on-write paging
which allows efficient forking; our Athlon64 3000+ running Debian
GNU/Linux can perform over 7,000 $fork()$ operations per second.

Since our model checker operates on programs with two threads, we
carefully synchronize pairs of processes to implement the possible
execution orders.  We call this technique ``the buddy system'', and a
pair of cooperating threads ``buddies''.  Each pair of buddies
maintains a data structure in shared memory containing semaphores,
execution path and other elements required for IPC and coordination.

And since $fork()$ creates processes, not threads, we implement
threadlike behavior using shared memory.  Threads do not get separate
copies of program variables as processes do, so we create a structure
in a shared memory segment where buddy processes keep all application
variables.

\subsection{Example}

We will expand the earlier example into separate statements to show
how our technique works.  Thread 0 executes $print(a);~print(b)$,
while thread 1 executes $print(1);~print(2)$.  First, we instrument
our code by placing calls to a function $hook()$ before each
statement, then calling $done()$ at the end of execution:

\begin{verbatim}

thread0() {
  hook(); print(a);
  hook(); print(b);
  done();
}

thread1() {
  hook(); print(1);
  hook(); print(2);
  done();
}

\end{verbatim}

The code is then compiled and linked with our model checking code.
When it executes, $fork()$ is called to create a separate process for
each thread, each of which executes into the first $hook()$.  There
are only two possible ways in which the threads can execute their
first statements; either $thread0$ or $thread1$ goes first.  Each
``thread'' (really a process) thus forks into two processes, resulting
in two parents and two children.  The pair of parents become buddies,
and the pair of children become buddies.  The parent processes each
wait for their children to terminate, much as a recursive function
calls itself and waits for the recursive call to finish.  The child
process for thread 1 blocks using a semaphore, waiting for its buddy
in thread 0 to execute a single statement.  The buddy process does
this by returning from the call to $hook()$, which allows the first
statement, $print(a)$, to execute.  Then that process hits the
next call to $hook()$ and signals to its buddy that it has finished
executing a statement.

Now the process repeats; either thread0 can execute another statement,
or thread1 can execute its first statement.  Again, each buddy forks,
with the parents waiting for the children to finish.  The child
process for thread 0 again goes first, returning to execute
$print(b)$, then calling $done()$, which signals to the buddy
process that thread 0 has completed execution.  With no remaining
alternatives, thread 1 now runs to completion, giving a resulting
output of ``ab12''.

Once the grandchildren of the original two threads have each
terminated, the children continue running.  Since the grandchildren
explored the case in which thread 0 executed another statement, the
children explore what happens when thread 1 runs.  Thread 1 returns
from its hook and executes $print(1)$, then signals its buddy.
Once again, the children have two alternatives, so they fork another
pair of grandchildren.  The thread 0 grandchild executes its second
statement and terminates, allowing thread 1 to complete and producing
the string ``a1b2''.  The children again execute a statement from
thread 1, $print(2)$, after which thread 0 runs to completion,
producing ``a12b''.  Now the original two threads can continue,
executing $print(1)$ from thread 1, and forking another set of
children, which fork grandchildren as before, producing the strings
``1ab2'', ``1a2b'', and ``12ab''.

\section{Code Instrumentation, Language Independence and Statement Atomicity}
\label{instrumentation}

In order for our system to work properly, application code must be
properly instrumented, by making calls to $hook()$ before each program
statement.  These statements are assumed to execute atomically, which
does not generally happen in current systems, but which can be assured
using a technique we describe later in this section.

This instrumentation process is very simple, and works independently
of language constructions like loops, function calls and branches.
For example, we first implemented the example we gave in the last
section as follows, essentially the same as we listed it before:

\begin{verbatim}
  if(child) {
    // thread 0
    hook(); printf(``a'');
    hook(); printf(``b'');
    done();

  } else {
    // thread 1

    hook(); printf(``1'');
    hook(); printf(``2'');
    done();
  }
\end{verbatim}

But later, we generalized it to work for arbitrary strings using a
separate function and a loop:

\begin{verbatim}
void str(char *s) {
  int i;

  for(i=0; s[i]; i++) {
    hook();  b->common.outstr[b->common.outidx++] = s[i];
  }
}

main() {
...
  if(child) {
    // thread 0
    str(``ab'');
    done();

  } else {
    // thread 1
    str(``12'');
    done();
  }
}
\end{verbatim}

A naive, automated instrumentation tool might have added additional
$hook()$ calls as follows, but that would have merely added overhead
to the model checking process:

\begin{verbatim}
void str(char *s) {
  int i;

  hook(); for(i=0; s[i]; i++) {
    hook();  b->common.outstr[b->common.outidx++] = s[i];
  }
}

main() {
...
  if(child) {
    // thread 0
    hook(); str(``ab'');
    done();

  } else {
    // thread 1
    hook(); str(``12'');
    done();
  }
}
\end{verbatim}

While modifying program source code before model checking is generally
deprecated, we feel our technique has several interesting features.
First, adding calls to $hook()$ before each program statement is easy
to do automatically for reasonably written source code, even without
constructing a formal parser for the target language.  The implementor
must simply avoid placing calls where they would cause syntax errors
in the program, such as in between function declarations.  Redundant
calls to $hook()$ add overhead, but don't otherwise break our system.
This makes our system straightforward to implement in a variety of
languages, whereas traditional systems require significant adaptation
to target languages in order to properly model their behavior.

Second, our instrumentation can be used to perform other tasks beside
model checking, by changing the behavior of $hook()$.  For instance,
$hook()$ could be modified to implement white box testing, in which
test cases are constructed which together must execute all code
branches.

Third, our instrumentation can be left in place to guarantee the
statement-level atomicity assumed by our system.  Generally speaking,
modern CPUs offer only machine instruction level atomicity -- the OS
may interrupt a process between any two instructions and begin
execution of a different process.  Model checkers like Estes\cite{estes} work on
these machine instructions directly, but their results can only be
applied to that particular compilation of the application on a
particular CPU.  This may prove to be the only way to prove useful
thread safety and liveness properties about unmodified code on a
particular CPU, and would tend to suggest that dealing with
multithreaded applications in their original high-level language form
doesn't even make sense.  On the other hand, if calls to $hook()$ are
left in place in distributed code, the function can be modified to
essentially make each application statement into an individual
critical section.

Admittedly, this adds a large amount of overhead to the code, since
system calls to raise and lower a semaphore must be made for each
program statement.  But in modern high level scripting languages
particularly, programs tend to have fewer statements, with powerful
built-in commands having relatively high execution costs.  Such
languages may be particularly difficult to verify at the machine code
level, since they run via large, complex interpreters.  As a first
approximation of the overhead our technique would add, we wrote a C
program which forks into two processes, each of which loops 1,000,000
times.  In the loop, each process grabs a semaphore, adds the current
index to a variable, then releases the semaphore.  The program
performed the 4,000,000 semaphore operations in about 0.8 seconds on
our Athlon64 3000+.  We then ran a program in Perl which loops
2,000,000 times in a single process, likewise adding up the index
values.  It also took about 0.8 seconds, suggesting that efficient
C-based hooks in the perl code to ensure statement-level atomicity
might add only 50\% overhead to such a program, and possibly less for
a program using fewer, more costly operations than simple loops and
additions.

\section{Formal Definitions}

Here we define terminology used in the algorithms described in the
next sections.

\begin{itemize}
\item Assume there exist functions $hook()$, which is called before
execution of each application thread program statement, and $done()$,
which is called after the last statement in each thread.  Application
threads may include most usual language features, such as branching,
looping and function calls (see section \ref{instrumentation}).

\item We define the {\bf execution counter} for a thread to be the
number of times $hook()$ has been called since the beginning of the
thread's execution.  Intuitively, this is the number of statements
executed in that thread, plus one.

\item If $s_0$ is the value of the execution counter for thread 0 and
$s_1$ is the corresponding counter for thread 1, the pair $\langle
s_0, s_1 \rangle$ forms the {\bf combined execution counter} for the
two threads.

\item An {\bf execution trace} is defined to be a string $t \in
\{0,1\}^*$ which represents the order in which statements from the two
threads were executed to reach a particular combined execution
counter.  In our earlier example, the execution trace $0011$
corresponds to the output ``ab12''.

\item Let $V_0$ and $V_1$ represent the vectors of shared variable
values for threads 0 and 1 having an execution trace $t$ and combined
execution counter $C$.  The tuple $I=\langle V_0, V_1, t, C \rangle$
is a {\bf partial interleaving} for the two threads (partial, since
the threads may not yet have run to completion).
\end{itemize}

\section{Search Algorithm}

Our algorithm for performing a depth first search on all possible
execution interleavings of two threads $t0$ and $t1$ is as follows:

\begin{itemize}

\item {\bf Base case:} Let $I$ be the initial partial interleaving for
  threads $t0,t1$, representing the program state at the first call to
  $hook()$ in each thread, before any application statements have
  executed.

\item {\bf Recursion:} Given a partial interleaving $I$ for threads $t0,t1$,
\begin{itemize}
  \item Run any user-supplied code for checking conditions.

  \item If $done()$ has been called in both threads, terminate the
  current thread and indicate successful program execution for a
  single complete interleaving.
  
  \item If $done()$ has been called in only one thread, allow the
  other thread to continue to completion.

  \item Otherwise, fork both threads to create children $t0',t1'$.
  Parents both wait for termination of the children.  $t0'$ returns
  from $hook()$, allowing a single statement to execute while $t1'$
  blocks.  Then the recursive step is performed again on $t0',t1'$
  with a new partial interleaving $I'$.  When $t0',t1'$ have
  terminated, $t1$ returns, executing a single statement, and then the
  recursive step is performed again for $t0,t1$ with new partial
  interleaving $I''$.
\end{itemize}
\end{itemize}

Omitting $\#include$ statements and helper functions for setting up
semaphores and shared memory, our C implementation of this algorithm
fits in one printed page of code (80 columns by 65 lines).

\section{Detecting Race Conditions and Pruning the Search Space}
\label{race}

Note that in the above example, program output is entirely dependant
on the order in which the OS schedules the two threads for execution.
Such nondeterministic behavior is almost never intended, and usually
represents a bug in the code.

Consequently, we provide a technique for verifying that threads behave
the same regardless of the order in which they are executed.  This
technique also makes it easy to avoid unnecessary exploration of the
space of possible interleavings.  Formally, we define a race condition
as follows:

\begin{itemize}
\item Let $I=\langle V_0, V_1, t, C \rangle$ and $I'=\langle V_0',
V_1', t', C' \rangle$ represent partial interleavings such that
$C=C'$.  That is, $I$ and $I'$ are partial interleavings which have
reached the same execution counter in each thread but potentially
through a different order of execution.  $I$ and $I'$ form a {\bf race
condition} iff $I \neq I'$.
\end{itemize}

To implement this technique, we maintain a shared table keyed on the
combined execution counters explored while searching the state space.
Each table entry records the partial interleaving at that combined
execution counter.  When a particular combined execution counter is
reached via a different execution trace, the current partial
interleaving is compared against the stored partial interleaving.  If
they differ, the two partial interleavings are displayed as examples
of execution paths capable of producing differing behaviors.  A
program with no race conditions will of course display only the single
possible outcome of program execution.

The second purpose of this table is to record which combined execution
counters have been reached before.  Since our algorithm performs a
depth first search on the possible thread interleavings, the second
occurrance of a combined execution counter can only occur once all the
remaining interleavings from that point on have already been explored.
Consequently, if a race condition does not occur at a particular
explored partial interleaving, there is no need to explore it again
since the two threads are in exactly the same state as they were the
last time.

Although we have not yet been able to derive a formula for the
complexity of our pruning algorithm, it is clear that this pruning
technique is at least an order of magnitude improvement over an
exhaustive search.  We ran our algorithm on pairs of threads each
executing 3 to 8 statements.  Values represent the total number of
calls to $hook()$, which roughly corresponds to the number of states
explored.

\begin{tabular}{lrrrrrr}
{\bf Technique}  & {\bf 3} & {\bf 4} & {\bf 5} & {\bf 6} & {\bf 7} & {\bf 8} \\
Exhaustive & 30 & 112 & 420 & 1584 & 6006 & 22876 \\
Pruning    & 18 &  32 &  50 & 72   &   98 &   128
\end{tabular}

This addition was surprisingly easy to add to our system; it required
less than a page of code in changes.

\section{Supporting IPC}

Our implementation supports multiple semaphores which may be used by
the user application for interprocess communication.  This complicates
our system, since a thread may block until the other thread releases a
particular semaphore, and complicates the actual implementation even
more, due to practical issues regarding process cleanup, IPC and
resource management.  Here we give the algorithm for supporting an
arbitrary number of application semaphores.

\begin{itemize}

\item Let the definition of a partial interleaving be extended to
  include a set of semaphores $S = \{s_0..s_n\}$, which may be up or down.

\item Let the function $down(i)$ be a valid application statement
  (to be preceded by a call to $hook()$).  $down(i)$ causes the current
  thread to lower $s_i$ if it is up, and do nothing otherwise.

\item Let the function $up(i)$ perform the complimentary
  operation, with the addition that if $s_i$ is already up, the thread
  blocks until the other thread calls $down(i)$.  If the other thread
  is already blocking, report deadlock and terminate both threads.

\item {\bf Base case:} Let $I$ be the initial partial interleaving for
  threads $t0,t1$, representing the program state at the first call to
  $hook()$ in each thread, before any application statements have
  executed.

\item {\bf Recursion:} Given a partial interleaving $I$ for threads $t0,t1$,
\begin{itemize}
  \item Run any user-supplied code for checking conditions.

  \item If $done()$ has been called in both threads, terminate the
  current thread and indicate successful program execution for a
  single complete interleaving.
  
  \item If $done()$ has been called in only one thread, allow the
  other thread to continue to completion.  If the other thread is
  blocked, report an error, since the thread will block forever.

  \item If a thread is blocked, allow the other thread to execute
  another statement.

  \item Otherwise, fork both threads to create children $t0',t1'$.
  Parents both wait for termination of the children.  $t0'$ returns
  from $hook()$, allowing a single statement to execute while $t1'$
  blocks.  Then the recursive step is performed again on $t0',t1'$
  with a new partial interleaving $I'$.  When $t0',t1'$ have
  terminated, $t1$ returns, executing a single statement, and then the
  recursive step is performed again for $t0,t1$ with new partial
  interleaving $I''$.
\end{itemize}
\end{itemize}

\section{Limitations}

In most thread implementations, threads share all global variables,
but each has its own stack.  Local variables are thus maintained
independantly from other threads.  Our implementation presently
provides no support for such variables, and assumes that all thread
state can be monitored via the shared variables and execution counter.
It would be easy to create a second structure associated
with each buddy process for storing variables unique to each thread,
and account for that additional state information when checking for
race conditions and pruning the state space.

As we described in section \ref{instrumentation}, our assumption that
program statements are performed atomically is not at all guaranteed
by real computers, unless our technique is employed at a machine code
level.  To achieve reported results in practice, statement-level
atomicity would need to be enforced by the operating system, language
interpreter, or by using a modified $hook()$ as we described.

While using $fork()$ to store program execution state makes our system
very simple to implement, it imposes a significant amount of system
overhead.  The system resources for two processes, including two
process table entries, are required for each level of depth in the
search space, which corresponds to the number of statements executed
by the combined threads.  This makes even simple programs, like a pair
of threads which each loops 1,000,000 times, impossible to verify with
our system.

Finally, our current system is limited to programs with two threads.
See section \ref{future} for discussion on removing this limitation.

\section{Implementation and Performance}

As we showed in section \ref{race}, our pruning algorithm performs far
better than an exhaustive search.  Default (though 
modifiable) OS limitations on the number of available semaphore sets
and our implementation's inefficient use of single semaphore sets
rather than multiple semaphores in multiple sets limits us to running
applications which execute a total of about 22 steps.  These it
handles in under 0.1 seconds.  The resulting maximum of 44 live
processes imposes no noticeable memory consumption.


While our fundamental search algorithm can be implemented in about a
page of code, our full system supporting application semaphores, race
detection and pruning, and with helper functions, debugging code, and
whitespace currently weighs in at 611 lines of C.  The implementation
requires 4 semaphores for each level of depth in the DFS, and requires
$n^3$ storage in the DFS depth to maintain the table of partial
interleavings complete with program execution paths.  For our current
depth limit of 22, this amounts to about 100k of memory.

\section{Future Work}
\label{future}

Our system is limited to programs with two threads.  Since our system
is modeled after the traditional recursive depth first search
algorithm, there is a clear path for extending our algorithm to
support any number of threads.  Rather than pairs of parent processes
spawning a single pair of children, each parent in a set of n parent
buddies would iteratively spawn n-1 children (waiting each time for
the previous child to die).  The n n-member buddy sets (cliques?)
would then be dispatched, exploring the paths in which each child
executes its next statement.  Instead of a maximum $2k$ live processes
for a k-level DFS, up to $nk$ processes would exist.

Implementation, in our experience, might be time consuming,
due to the inherent difficulty humans seem to have keeping track of
multiprocess systems.  However, with careful planning, and for
programmers more experienced with multiprocess applications, this
extension should not prove too difficult.

One feature that might be quite easy to add is the ability for parents
to run without waiting for their children.  Unchecked, this would act
like a ``fork bomb'', potentially swamping the system as the entire
search tree unfolded at once.  But with a limit on how many pairs of
processes could run at once, our system would immediately be able to
run on SMP systems with arbitrary numbers of processors, limited only
by the size of the process table and the system memory.

At the other extreme, with only two processes running at once, it's
unnecessary to allocate each pair of buddies their own set of
semaphores, as our first implementation does.  
With more careful use and management, we suspect that a
single set would suffice, avoiding system limits on available
semaphores.  The state table also need not require so much storage;
cryptographic hash functions can be used to reduce arbitrary amounts
of data to a 128 bit digest which can be stored in place of the actual
partial interleaving.  An expected $2^{64}$ entries would have to be
made before any pair of entries would have the same digest, which is
far more than any PC can store today.

As we described in section \ref{instrumentation}, our system should be
easy to implement in a variety of languages, possibly just by using
cross-language extensions to call our original C implementation.
There are also other uses for $hook()$ which we described but have not
implemented.

\section{Conclusion}

We gave a straightforward technique for checking thread-related
properties of programs with two threads.  Our technique is
general-purpose and language-independent, and may be particularly
suited to modern high-level scripting languages due to the difficulty
of machine-code level model checking for these languages and the
overhead required to enforce the statement-level atomicity required by
our system.

Our system performs much better than an exhaustive search of the state
space, understands application semaphore use, and detects deadlock and
race conditions.  Furthermore, our implementation is quite compact;
our exhaustive search algorithm fits in one printed page of C code,
and our complete implementation fits in under ten.

\end{document}